# Supporting Error Chains in Static Analysis for Precise Evaluation Results and Enhanced Usability

Anna-Katharina Wickert [*], Michael Schlichtig [†], Marvin Vogel [‡],
Lukas Winter [§], Mira Mezini [*,¶], Eric Bodden [†]

[*]Technische Universität Darmstadt
Darmstadt, Germany
lastname@cs.tu-darmstadt.de
[†]Heinz Nixdorf Institute
Paderborn University
Paderborn, Germany
firstname.lastname@uni-paderborn.de

[‡]Universität Hamburg
Hamburg, Germany
marvin.vogel@studium.uni-hamburg.de
[§]Unaffiliated
mail@lukas-winter.net
[¶]Hessian Center for Artificial Intelligence (hessian.AI)
National Research Center for Applied Cybersecurity ATHENE

*Abstract*—***Context:*** Static analyses are well-established to aid in understanding bugs or vulnerabilities during the development process or in large-scale studies. A low false-positive rate is essential for the adaption in practice and for precise results of empirical studies. Unfortunately, static analyses tend to report where a vulnerability manifests rather than the fix location. This can cause presumed false positives or imprecise results. ***Method:*** To address this problem, we designed an adaption of an existing static analysis algorithm that can distinguish between a manifestation and fix location, and reports error chains. An error chain represents at least two interconnected errors that occur successively, thus building the connection between the fix and manifestation location. We used our tool *CogniCrypt$_{SUBS}$* for a case study on 471 GitHub repositories, a performance benchmark to compare different analysis configurations, and conducted an expert interview. ***Result:*** We found that 50% of the projects with a report had at least one error chain. Our runtime benchmark demonstrated that our improvement caused only a minimal runtime overhead of less than 4%. The results of our expert interview indicate that with our adapted version participants require fewer executions of the analysis. ***Conclusion:*** Our results indicate that error chains occur frequently in real-world projects, and ignoring them can lead to imprecise evaluation results. The runtime benchmark indicates that our tool is a feasible and efficient solution for detecting error chains in real-world projects. Further, our results gave a hint that the usability of static analyses may benefit from supporting error chains.

## I. Introduction

Static analysis is a crucial tool that identifies software bugs and vulnerabilities, and is highly recommended by experts [1], [2]. For cryptography, the available tools range from command-line solutions [3], [4] over IDE integrations [5], and software-as-a-service solutions [6], [7]. Researchers and practitioners developed these tools to respond to the challenges of using cryptography securely [8], [9]. Because of their abilities, experts recommend integrating static analyses to detect security bugs in the development cycle [2]. This ultimately leads to more secure software for all users. However, previous work has shown that many false positives hinder the adaption of static analyses [10], [11]. Confusing reports may also render the analysis' result to be unresolved [12] or result into *effective false positives* [13], [14]. Further, analyses with many false positives may draw an imprecise picture of bugs or vulnerabilities in software when used in empirical studies.

In 2004, Avizienis et al. [15] discussed the basic concept of errors for security. In particular, they point out that one *fault* can cause another *error* and these errors can propagate. The propagation process transforms one *error* into multiple *errors* that can occur across different components. This behavior of an API usage is also known as *multi-object protocol* which is prevalent in cryptographic APIs [16], [17]. Further, Lipp et al. [18] revealed that static analyses rather report the *manifestation* of an error than its *root* cause (*fault*). They argue that this decision improves the precision of analyses, as not every *fault* manifests as a *vulnerability*. An empirical study [14] revealed that 20 % of the analyzed cryptographic misuses could not be resolved, as the *root* cause did not match the reported location. Concretely, while a misuse was reported, the required fix was unclear from the report. Summarizing, especially in cryptography, errors can propagate. Therefore, an error can depend on other errors, thus resulting in error chains. Such a *dependent error* in an error chain is a *subsequent error* to its *preceding errors* of the error chain.

Typically, studies and tools do not address error chains. In this paper, we aim to close this research gap. We focus on the domain of cryptographic API misuses in Java as a well-researched area with many detection tools that cannot (yet) provide this granularity of information. Concretely, we want to understand to which extent error chains occur in the wild, if analyses that present the complete error chain are usable in practice concerning the runtime, and if the presentation of an error chain can improve the usability of a security-focused static analysis. To achieve this, we will introduce *CogniCrypt$_{SUBS}$* which is our adaption of an existing static analysis to allow the detection of error chains. We will answer the following three novel research questions:

- [RQ1] What is the distribution of dependent errors for real-world applications?
- [RQ2] What is the runtime overhead of *CogniCrypt$_{SUBS}$* compared to an error chain unaware version of *CogniCrypt$_{SAST}$*?
- [RQ3] To which extent do experts consider *CogniCrypt$_{SUBS}$* as usable compared to an error chain unaware version of *CogniCrypt$_{SAST}$*?

By answering these research questions, we aim to provide new insights into error chains in practice. Understanding error chains for cryptographic API misuses in Java can support us in understanding empirical studies in more depth. In particular, the insights can help us to understand the real-world prevalence of error chains. Further, we gain new insights into the challenges of expressing the dependency between a root error and its subsequent errors for static analyses. In addition, our expert interview sheds light on the possible usability implications of reporting error chains.

To address this problem, we designed the first adaption of an existing static analysis algorithm to distinguish between root and subsequent errors. We adapted the algorithm of an existing and established security-focused static analysis, namely *CogniCrypt$_{SAST}$*[3]. *CogniCrypt$_{SAST}$* was used in several empirical studies [3], [14], [19], [20] and can be used to analyze further APIs beyond cryptography. Further, we quantitatively evaluated real-world repositories with our adaption, called *CogniCrypt$_{SUBS}$*, measured the runtime overhead, and conducted a user study.

The empirical evaluation for *RQ1* revealed that at least half of the projects with at least one cryptographic API contain error chains. The runtime evaluation for *RQ2* showed that our adaptions lead to an on average overhead between 1%—4%. The expert interview for *RQ3* implicates that *CogniCrypt$_{SUBS}$* helped the participants to identify the error chain more quickly and reduced the required interaction time with the tool.

Overall, this paper presents the following contributions:

⋆ We present an algorithm for static analyses that can distinguish between a root and subsequent errors.
⋆ We improve an existing open-source analysis to integrate the suggested changes.
⋆ We show that error chains occur in the wild and affect every second analyzed project.
⋆ We present a runtime evaluation that reveals that the overall median runtime overhead of the more detailed reporting is 3.78 %.
⋆ We report on an expert interview that revealed that the participants appreciate the structured reporting of root and subsequent errors and require fewer executions of the analysis to resolve all potential problems.

In the remainder of the paper, we first introduce the basic concepts of the analysis that we extend in Section II, followed by a description of the terminology to present error chains. We describe the required adaptions to an existing algorithm in Section III. Section IV briefly describes the motivation of our research questions, methodology, and findings. We cover related work in Section V and limitations and future work in Section VI. The paper ends with a conclusion in Section VII and information about our artifacts in Section VIII.

## II. FUNDAMENTALS OF *CogniCrypt$_{SAST}$*

To implement our static analysis adaption *CogniCrypt$_{SUBS}$* to detect error chains, we extend the static analysis *CogniCrypt$_{SAST}$* [3], [21]. Thus, this section introduces the necessary information to understand our approach (c.f. Section III-B). First, we introduce an example of a cryptographic misuse (Section II-A), then explain the different error types reported by *CogniCrypt$_{SAST}$* (Section II-B), and third, we shortly describe the analysis of *CogniCrypt$_{SAST}$* (Section II-C).

### A. Cryptographic Misuse Example

In Listing 1, we present an example of an insecure file encryption in Java. Our example has two methods, `generateKey` (Line 2) and `encryptFile` (Line 8). The method `generateKey` generates a secret key using the JCA classes `DESKeySpec` and `SecretKeyFactory` from the given password. This method is called by the method `encryptFile` in Line 11. The method `encryptFile` generates an initialization vector (Line 9), generates and initializes a `Cipher` object in Line 10 and 11 respectively, and passes the `Cipher` object to a `CipherInputStream` in Line 13.

Listing 1 exhibits seven different cryptographic misuses detected by *CogniCrypt$_{SAST}$*. The first and second misuses on Lines 4 and 5 respectively, occur because the insecure algorithm DES is passed to the object `keyFactory` and the class `DESKeySpec` is used. These two misuses result in another report for the second parameter of the `Cipher.init` method in Line 11, which results in an insecure cipher input stream object in Line 13. Further, the misuse in Line 9 is due to the static initialization vector, causing another misuse for the third argument of the `Cipher.init` call in Line 11. Finally, the `Cipher` object is initialized with an insecure algorithm in Line 10 that also renders the `Cipher` object insecure.

### B. CrySl and Error Types

*CogniCrypt$_{SAST}$* uses an allow-listing approach, where rules describe the secure usage of cryptographic APIs in the domain-specific language *CrySL* [3]. Currently, *CrySL* supports classes of the Java Cryptography Architecture (JCA) and Java Secure Socket Extension (JSSE) APIs. Violating a rule causes one of two main error types: *typestate errors* and *constraint errors*.

A rule can express the secure order of method calls. If there exists one typestate path for a *specified object*, an instance of a class for which a *CrySL* rule exists, that violates the

```
1   public class FileEncrypt {
2     private static Key generateKey(String password) throws Exception {
3       DESKeySpec dks = new DESKeySpec(password.getBytes("utf8"));
4  🐛    SecretKeyFactory keyFactory = SecretKeyFactory.getInstance("DES");
5  🐛    return keyFactory.generateSecret(dks);
6     }
7
8     public static String encryptFile(String password, String srcFile, String destFile) {
9  🐛    IvParameterSpec iv = new IvParameterSpec("123456".getBytes("utf-8"));
10 🐛    Cipher cipher = Cipher.getInstance("DES/CBC/PKCS5Padding");
11 🐛    cipher.init(Cipher.ENCRYPT_MODE, generateKey(password), iv);
12      InputStream is = new FileInputStream(srcFile);
13 🐛    CipherInputStream cis = new CipherInputStream(is, cipher);
14      // write cis into destFile and return
15    }
16  }
```

Listing 1. A simplified example of an insecure encryption observed in the wild. 🐛 represents an API misuse.

specified order, *CogniCrypt$_{SAST}$* reports a *typestate error*. A *typestate error* can be an `IncompleteOperationError` if the secure order is not completed or a `TypestateError` for any other derivation of the secure call order.

Each rule can express constraints. A *constraint error* describes a violation of a requirement expressed by a rule, such as a secure encryption algorithm. Overall, five different *constraint errors* exist. A `ConstraintError` is reported if a constraint on any of the objects specified in the *CrySL* rule is violated. Some *specified objects* require a predicate that signifies a guarantee from another API, i.e., a predicate describes interactions between classes. The *specified object* that can ensure a predicate is an *ensuring object*. If a *specified object* requires a predicate that is not ensured by an *ensuring object*, this results in a `RequiredPredicateError`. If sensitive hard-coded information, insecure types, or forbidden methods were used, *CogniCrypt$_{SAST}$* reports a `HardCoded-Error`, `NeverTypeOfError`, or `ForbiddenMethod-Error`, respectively.

Listing 1 illustrates the concept of *ensuring* and *specified objects*. The rule for the JCA class `CipherInputStream`[1] requires that the argument (Line 13) of the class `Cipher` is generated securely. Thus, the object `cis` requires a security guarantee from the class `Cipher`. The object `cis` is the *specified object* and the object `cipher` is the *ensuring object*. As the *ensuring object* is insecure due to the use of DES[2] as encryption algorithm (Line 10), the predicate is not ensured, resulting in a `RequiredPredicateError` (Line 13).

### C. Analysis Algorithm

The static analysis *CogniCrypt$_{SAST}$* uses allow-listing rules to identify misuses [3], [21] (c.f. Section II-B). We present a shortened and simplified version of the (final) algorithm in Listing 2. The analysis first parses the rules and builds a finite state machine (FSM) for each rule (Line 18). Each FSM encodes the call order for one class that is encoded in the respective *CrySL* rule. Afterward, the analysis generates a call graph with the analysis framework *Soot* [22] (Line 19). This call graph serves as a starting point to identify allocation sites that match the initial edge of all FSMs (Line 21).

[1]The respective *CrySL* rule is available online and the discussed requirement is specified in Line 30 and 31 (tag 3.0.2-jca): https://github.com/CROSSINGTUD/Crypto-API-Rules/blob/master/JavaCryptographicArchitecture/src/CipherInputStream.crysl.

[2]The respective *CrySL* rule is available online and the secure values for the argument are specified in Line 88 to 118 (tag 3.0.2-jca): https://github.com/CROSSINGTUD/Crypto-API-Rules/blob/master/JavaCryptographicArchitecture/src/Cipher.crysl

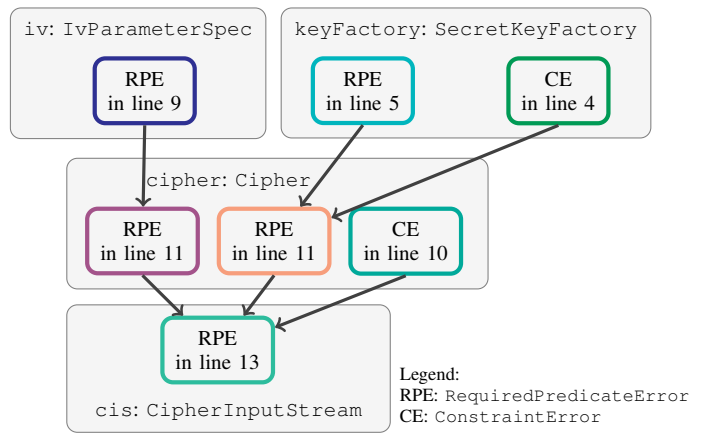

Fig. 1. Dependent error tree of an error reported for `CipherInputStream` in the code of Listing 1. The gray boxes mark different objects and the differnt box colors help to differentiate misuse locations.

For each identified seed, *CogniCrypt$_{SAST}$* uses *IDE$^{al}$* [23] and the matching FSM to perform a flow-, field-, and context-sensitive typestate analysis. To retrieve the parameter values of methods that are invoked on the seed object, *CogniCrypt$_{SAST}$* executes a backward pointer analysis [24]. Both the results of the typestate and backward analysis are stored in the respective seed. A seed uses this information to report errors in three stages. First, the results of the typestate analysis are used to report *typestate errors* (c.f. Section II-B, Line 26). Second, a seed validates the specified constraints and reports *constraint errors* (c.f. Section II-B, Line 27). Third, a seed triggers

```
17  // Initialize analysis
18  fsms := parseRules(cryslRules)
19  cfg := generateCallGraph(program)
20  // Pre-analysis to identify crypto objects
21  for edge.allocation(fsms.classes) in cfg {
22    objects.add(edge)
23  }
24  // Analyze crypto objects traces
25  for seed in objects {
26    identifyTypeStateErrors(seed)
27    identifyConstraintErrors(seed)
28  }
29  // Propagate until all seeds updated
30  for handler.hasUpdates() {
31    for p in handler.updatedSeeds().preds() {
32      if p.noViolatedConstraints() &&
         ↪  p.requiredPredicateError() {
33        ensuredPredicate(p)
34      } else {
35        hiddenPredicate(p)
36      }
37      handler.passToUsingObjects(p)
38    }
39    // Handle and map subsequent errors
40    for rpe in requiredPredicateErrors {
41      if rpe.isSubsequent() {
42        precedingErrs := rpe.getHiddenPred()
43        identifyMatching(precedingErrs)
44      }
45    }
46  }
```

Listing 2. Simplified Pseudo-code of the Analysis Algorithm

the propagation of security guarantees, such as that a key is generated securely (Line 25).

The aim of the propagation is to track security guarantees across multiple objects. Each security guarantee is a *predicate*. If a seed has no violated constraint and all required predicates (Line 32), such as a salt for a password-based encryption must be randomized securely, are ensured, it ensures its predicate (Line 33). The *ensured predicate* is passed to a handler that passes the newly *ensured predicate* further to all objects with a *CrySL* rule that use the object of the *ensured predicate* (Line 37). The propagation repeats when a seed receives a new *ensured predicate* and terminates when all seeds finished ensuring any predicate (Line 25).

## III. DESIGN

In this section, we first discuss relevant terminology (Section III-A). We then explain our approach and necessary algorithm adaptations to detect error chains (Section III-B).

### A. Terminology

To securely encrypt the file, the code must be modified at four locations, and yet, analyses, such as *CogniCrypt$_{SAST}$*, tend to report up to seven misuses (c.f. Listing 1). However, the misuses actually form a tree of misuses (c.f. Figure 1). We refer to errors with preceding errors as *subsequent errors*. A *subsequent error* is caused by another misuse that is the respective *preceding error*. For instance, the misuse reported in Line 11 is a *subsequent error* caused by the insecure initialization vector from the *preceding error* in Line 9. If a developer fixes a *preceding error*, the *subsequent error* is fixed, too (if all *preceding errors* are resolved). As a *preceding error* can be a *subsequent error* too, we use the term *root error* to describe any *preceding error* that is not a *subsequent error*. In Figure 1, the ConstraintError in the right top (Line 4) is a *root error* and is a directly *preceding error* for the RequiredPredicateError in the middle (Line 11). This error is the *subsequent error* (Line 11) of the *root error* (Line 4) and it as well is a *preceding error* of the RequiredPredicateError (Line 13) at the bottom of Figure 1. This error, is the *subsequent error* (Line 13) of the two *preceding errors* (Line 4 and 11). The current implementation of *CogniCrypt$_{SAST}$* — and other analyses in the domain — only reports errors unsorted without reporting their connection. Thus, the report may be confusing or include *effective false positives* [12], [13], [14], and ultimately may hinder the adaption of such tools [10], [11].

### B. Hidden Predicate Approach Implementation

In the discussed example in Listing 1 and Figure 1, we could see that error chains occur when passed arguments do not hold a required property. Such properties can be described as *predicates* (c.f. Section II-B). Since *CogniCrypt$_{SAST}$* already provides a mechanism to identify *predicates* (c.f. Section II-C), we decided to adapt its static analysis algorithm, and we call it the *Hidden Predicate Approach*.

> **Hidden Predicate**
>
> A *hidden predicate* is generated if an *ensuring object* fails to validate all its specifications (i.e., no *predicate* was generated).

Our approach is based on the concept of tracking *hidden predicates* – whenever a predicate failed to be ensured, we collect a *hidden predicate* instead of an *ensured predicate* (Line 35). The main design decision for our implementation is that *hidden predicates* should be propagated in the same way as ensured predicates. Consequently, we enriched the existing propagation mechanism of *CogniCrypt$_{SAST}$* to integrate the *hidden predicate* propagation. Our adaption (Listing 2) of the *CogniCrypt$_{SAST}$* algorithm (c.f. Section II-C) has three main changes: the generation of *hidden predicates* (Line 35), the detection of *subsequent errors* (Line 41), and the mapping of *subsequent* and *preceding errors* (Line 42 and 43).

To ensure that the algorithm propagates *hidden predicates* as *ensured predicates*, we handle *hidden predicates* as a subclass of *ensured predicates*. A seed generates a predicate either as ensured or hidden (Line 32). A *hidden predicate* is generated when a seed does not ensure a predicate (Line 35). The generated *hidden predicate* is passed further to any object that might require the *hidden predicate* to be ensured (Line 37). Therefore, a seed receives *hidden predicates* by the same logic as *ensured predicates*.

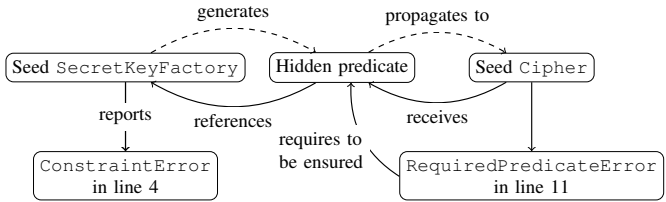

Fig. 2. Example reference between the `SecretKeyFactory` and `Cipher` seed for Listing 1.

To connect the different errors to generate an error chain, we rely on our hypothesis that subsequent errors are caused by unfulfilled predicates. Thus, the seed checks for each determined and reported *required predicate error* if this error is a subsequent error (Line 40 and 41). The error is subsequent if the seed received a hidden predicate (Line 42) that includes the reason for the *missing ensured predicate* (Line 43). Consider the example in Figure 2. The seed for `Cipher` has a `RequiredPredicateError`. Thus, the seed (Line 42) receives the *hidden predicate* that references the seed for `SecretKeyFactory`. As this seed has a `ConstraintError`, the algorithm can determine that this `ConstraintError` is the *root error* of the discussed `RequiredPredicateError`.

Note that our implementation of the *hidden predicate approach* enabled us to implement another improvement for a well-known over approximation causing false positives. For instance, a *Cipher* object uses a *IvParamaterSpec* object and requires a securely randomized byte array generated by an object of *SecureRandom* as IV only in the case of encryption and not for decryption. *CogniCrypt$_{SAST}$* falsely reports a `RequiredPredicateError` for a static IV for decryption and our change in *CogniCrypt$_{SUBS}$* and the *CrySL* rule set no longer reports this false positive. For this change, we leveraged an unused feature of *CrySL*'s grammar that can express implications, and adapted the *CrySL* rules to include the implications where required. Due to our changes with the *hidden predicate* approach, we can track the implications. We call this change that avoids instances of the above-mentioned false positives backward error tracking (BET).

## IV. EVALUATION

This Section, presents the results of our empirical study, benchmark, and expert interview. For each research question, we describe the motivation and methodology used to answer the research question and discuss the obtained results.

### A. Large-scale Study of JCA misuses in the wild (RQ1)

Previous work has pointed out that static analyses tend to report the location where a bug or vulnerability manifests, rather than where it is required to be fixed [18]. For Listing 1, consider the insecure initialization vector. The insecurity manifests in Line 11 and should be fixed in Line 9. For cryptographic misuses, previous work showed that a missing distinction between the dependency of errors can draw imprecise results of API misuses in the wild [14]. Further, this may hinder the adaption of static analyses for this critical field, as this may cause *effective false positives* [13], [14]. *Effective false positives* are true positives reported by the analysis that are perceived as false positives and not fixed by developers because, e.g., developers do not understand the report explanation [13]. Hence, the motivation of this research question is to understand the dependency between errors in the wild for cryptographic API misuses. In particular, we were interested if dependent errors are common in real-world projects, how errors can be depending on each other, and which cryptographic classes are most commonly affected.

*1) Methodology:* To analyze real-world projects, we mined popular projects on GitHub. We used a repository miner [3] to collect popular and active Java projects (c.f. Section I). Concretely, all projects should be Java projects, have at least 100 stars (popular), at least one commit between 02/22/2022 and 08/22/2022 (active at the time of mining), and be no fork (no duplicates). These filters resulted in 4,022 distinct repositories.

To analyze the projects with *CogniCrypt$_{SUBS}$*, we require binaries. Therefore, we first filtered for projects containing a JCA class, namely `java.security`, `javax.crypto`, or `java.*`, since our target domain is cryptography, and identified 2,054 projects that may use the JCA. Second, we tried to compile the remaining 2,054 projects automatically with Maven and Gradle with a time limit of 20 minutes. We succeeded for 634 repositories. To ensure that the projects use JCA, we executed a lightweight *OPAL* [25] analysis that identifies if any class of the JCA is used in the binaries. This step reduced the number of projects that we analyzed with *CogniCrypt$_{SUBS}$* to 505 repositories with in total 833 modules. We use the term modules to attribute the number of `bin` folders that we obtained through the compilation process.

With *CogniCrypt$_{SUBS}$*, we analyzed 505 Java repositories to identify error chains in the wild. We analyzed each repository with 3 GB memory, 2 MB stack size, and a timeout of 20 minutes for each project. With these restrictions, we received reports for 471 repositories and 783 modules. The repositories have 100 to 60,637 stars, and the median date of the latest commit was 15 days before we started the study.

*2) Findings:* To answer RQ1, we first focus on the quantitative results obtained through our analysis. Second, we discuss one large example as a more qualitative inspection.

*a) Quantitative Results:* *CogniCrypt$_{SUBS}$* reports at least one error for 306 of the 471 analyzed repositories. In total, *CogniCrypt$_{SUBS}$* reports 3,964 errors with on average 8.41 errors per repository, and 5.06 per module. `RequiredPredicateErrors`, `ConstraintErrors`, and `IncompleteOperationErrors` are the most common errors that *CogniCrypt$_{SUBS}$* observes in our dataset. Table I presents the number of errors for each of the different error types defined by *CogniCrypt$_{SAST}$*[3] (c.f. Section II-B). In our dataset, error reports for the JCA classes `Cipher`, `MessageDigest`, and `SSLContext` cause most errors.

[3]https://github.com/marvinvo/Repository_Miner

TABLE I
OVERVIEW OF THE NUMBER OF IDENTIFIED MISUSES.

| Error Type | Errors | Root Errors |
|---|---|---|
| RequiredPredicateError | 1617 | 353 |
| ConstraintError | 810 | 111 |
| IncompleteOperationError | 732 | 1 |
| TypestateError | 185 | 8 |
| HardCodedError | 183 | 127 |
| NeverTypeOfError | 176 | 117 |
| ForbiddenMethodError | 17 | - |

The usage of the insecure algorithm MD5 or SHA1 in `MessageDigest` is the most commonly reported misuse with in total 297 errors in 213 modules and 179 repositories.

In 155 (50.7%) repositories of the 306 repositories containing at least one error, *CogniCrypt$_{SUBS}$* reports at least one dependent error. In our dataset, 717 errors are marked as root errors and with 353 `RequiredPredicateErrors` are the most prevalent error type. Overall, only 9 root errors are typestate errors. Therefore, root errors caused by a *specified object* that does not reach a predicate-ensuring state are edge cases in our dataset. The remaining 98.7 % root errors are constraint errors (c. f. Section II-B). The number of root errors caused for each error type is presented in Table I.

In Listing 1, we showed that one subsequent error may have multiple preceding errors. The results of *CogniCrypt$_{SUBS}$* confirm this in practice. On average, each root error has 1.1 directly followed subsequent errors. Each subsequent error has on average 1.54 preceding errors. For 134 subsequent error trees the depth is equal or higher than 3. On average, a dependent error is a directed-acyclic graph with 3.83 errors, and the largest in our dataset contains 22 errors.

A few JCA and JSSE classes cause dependent errors. Overall, *CogniCrypt$_{SUBS}$* reported dependent error pairs for 24 distinct JCA classes. Most of the root errors are reported for the JCA classes `SecretKeySpec` (240), `KeyStore` (150), and `KeyManagerFactory` (96) that either serve as key store or wrapper. Thus, insecure keys cause most of the dependent errors. As a result, the three JCA classes with most of the subsequent errors require a key for their task. Namely, the JCA classes `Cipher`, `SSLContext`, and `Mac` with 219, 79, and 78 reported subsequent errors, respectively. Further, three JCA classes, namely `AlgorithmParameterGenerator`, `IvParameterSpec`, and `GCMParameterSpec` cause only root errors in our study. We present an overview of the number and dependencies between these classes in Figure 3. Note, we removed self-loops from the classes `SecureRandom` and `KeyPair` caused by requirements within the class. For example, an instance of `SecureRandom` can be initialized with a seed. This seed needs to be securely randomized, and this requirement can be ensured by another call sequence of `SecureRandom`.

> **RQ1**
> In-the-wild, every second project with cryptographic misuses has dependent errors.

*b) Case Study:* In this case study, we investigate one of the largest chains of subsequent errors found in our study. To calculate the length, we computed each error's preceding and subsequent error tree and appended the subsequent tree to the preceding error tree. This step results in a graph that we call the *dependent error tree* of an error. The case study discusses one of the errors with most edges in their dependent error tree.

We already discussed this case as an example in Figure 1 (c. f. Section III) which presents the dependent error tree for an insecure file encryption. Listing 1 presents a minified example resulting in this dependency tree. We observed this tree in a Java project with 699 stars, and the tree is caused by three mistakes to securely encrypt a file that we discussed in more depth in Section III-B. First, the initialization vector is insecure and resulting in three `RequiredPredicateErrors` as shown in the left part of Figure 1. Second, the key is insecure and causes the subtree that originates from the `RequiredPredicateError` and `ConstraintError` of the JCA class `SecretKeyFactory`. Third, the encryption algorithm is insecure and causes the `ConstraintError` in the `Cipher` object. Therefore, *CogniCrypt$_{SUBS}$* correctly reported this dependent error tree.

*B. Runtime Evaluation (RQ2)*

To identify subsequent and root errors, we adapted the algorithm to identify misuses to track more information. Thus, we assume that our changes will cause a runtime overhead. As the overall performance of an evaluation has an impact on their usability for empirical studies and use by developers, we were interested in the runtime impact of *CogniCrypt$_{SUBS}$*.

*1) Methodology:* To answer RQ2, we use a benchmark to compare the runtime of different configurations of *CogniCrypt$_{SAST}$* including our adaption of *CogniCrypt$_{SUBS}$*. Our initial assumption was that the overall number of errors in an application will affect the runtime differences between the different configurations. Specifically, we were more interested in the runtime overhead caused by our adaption of the algorithm than on the general runtime of *CogniCrypt$_{SAST}$*. As our motivation for the benchmarks started from a practical perspective, we decided to include real-world examples rather than synthetic test cases for our benchmark. Therefore, we used the dataset and results obtained in RQ1 to select modules that serve as a base for our benchmark.

To measure and compare the impact of the number of reported errors, we choose four different sets with five different modules each. The first, second, and third set includes modules with different number of errors observed in the dataset without a selection criteria focused on subsequent errors. Specifically, the first set has modules that have one error, the second set has modules with nine (observed average) errors, and the third set has the modules with the highest number of errors. Further, to focus on the impact of subsequent errors, the fourth set includes the five modules with most subsequent errors.

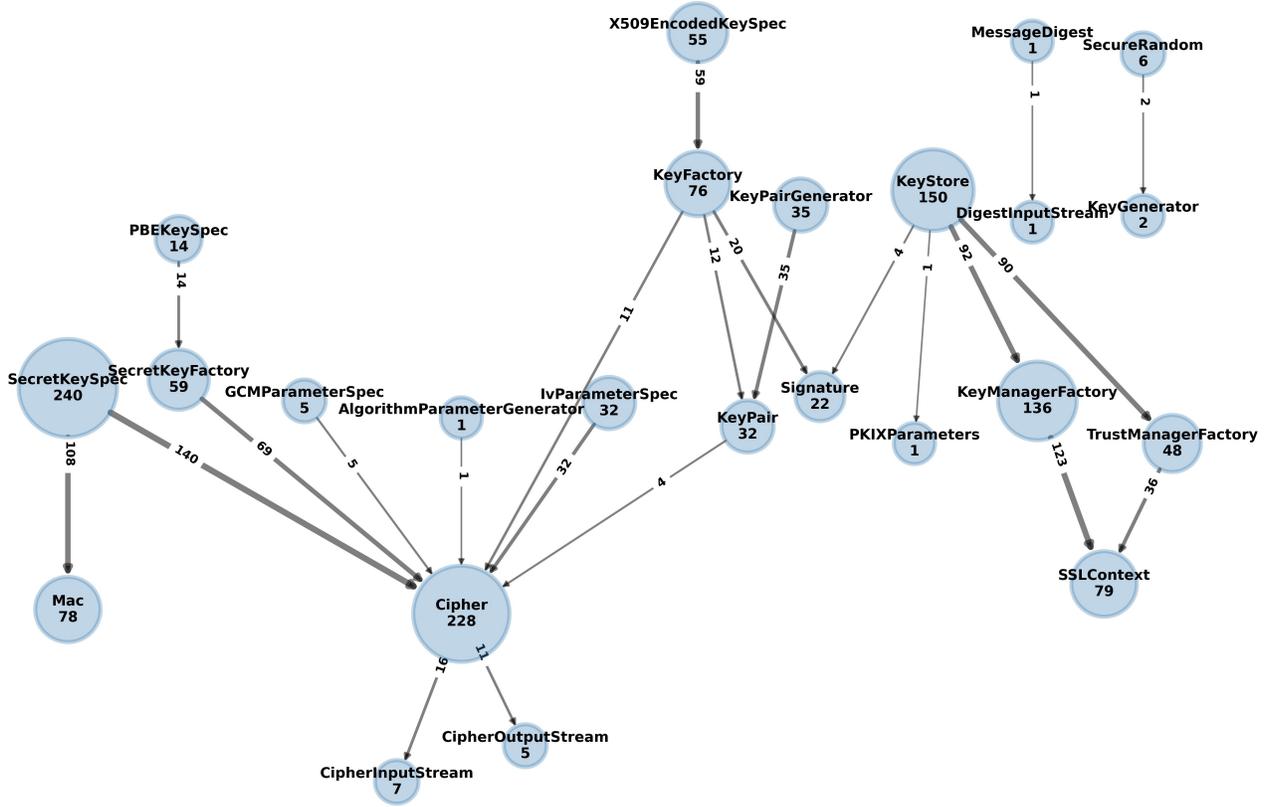

Fig. 3. JCA classes and their number of dependent errors. Each node presents the number of dependent errors for the JCA class. A directed edge shows the number of preceding errors that cause subsequent errors in another class.

We compared four different tool configurations (c. f. Table II). We used the version of *CogniCrypt$_{SUBS}$* that we used for our study in RQ1 (SUBS) including the BET and our adaptions to the rule set (Section III-B). To ensure fairness, all other three configurations used the latest (at the time of the study) public *CrySL* rule set [4] without BET. Note, we had to restructure and adapt the implementation of the *CogniCrypt$_{SAST}$* algorithm to implement the functionality that was required for *CogniCrypt$_{SUBS}$*. Therefore, our changes may have an impact on the runtime. As a baseline, we used the latest [5] published version of *CogniCrypt$_{SAST}$* (SAST).

We executed the benchmark within a Docker container on a 2-core machine with 32 GB RAM. For each analysis execution, we assigned 10 GB RAM and 10 MB stacksize. Each tool analyzed each module 10 times. For each measurement, we measured the elapsed time after selected analysis phases, such as the time in milliseconds to read the rules or collect and report missing predicates. For *CogniCrypt$_{SUBS}$*, we also

[4]commit 82d575190e, https://github.com/CROSSINGTUD/Crypto-API-Rules/tree/82d575190e/JavaCryptographicArchitecture
[5]commit 35d09163f, https://github.com/CROSSINGTUD/CryptoAnalysis/commit/35d09163f

TABLE II
DIFFERENT TOOL CONFIGURATIONS USED IN THE BENCHMARK.
(SED = SUBSEQUENT ERROR DETECTION, BET = BACKWARD ERROR TRACING, - = FEATURE NOT AVAILABLE)

| Acronym | Tool | Ruleset | SED | BET |
| --- | --- | --- | --- | --- |
| SUBS | CogniCryptSUBS | our improved | on | on |
| SUBSON | CogniCryptSUBS | latest public | on | - |
| SUBSOFF | CogniCryptSUBS | latest public | off | - |
| SAST | CogniCryptSAST | latest public | - | - |

measure the time required to detect and map subsequent errors.

*2) Findings:* We present the results of our runtime benchmark with a box-and-whisker plot. Each row visualizes all data points we collected for the specific measurement plotted. The box plot presents the 25 % to 75 % quartile and the whisker the variability outside the quartiles. The dots present a single measurement.

Figure 4[6] presents the overall runtime differences between all tools and configurations. Our finding is that for the whole workflow of the analysis, *CogniCrypt$_{SUBS}$* with activated subsequent error detection add a slight runtime overhead. On median the runtime overhead of subsequent error detection

[6]We added plots for other steps in our artifact (c. f. Section VIII).

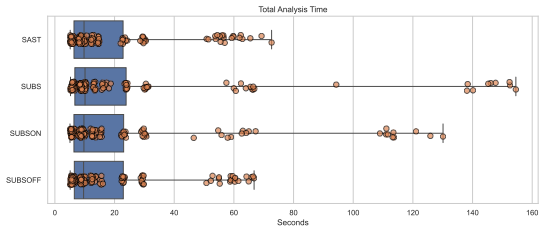

Fig. 4. Total analysis time of all tool configurations as presented in Table II

(*SUBSON*) is 1.16 % and for the configuration with BET (*SUBS*) is 3.78 %.

All tool configurations finished parsing the *CrySL* rules on average in 2 seconds. However, *SUBS* caused a small runtime overhead compared to the other configurations, which we attribute to the extended rule set. The overall analysis time of all tools and configurations confirms this observation, as presented in Figure 4. Further, *SUBS* has a small runtime overhead of 63 milliseconds compared to *SAST*, *SUBSON*, and *SUBSOFF*. We also attribute this overhead to the improved rule set that we only used in *SUBS* to enable *BET*.

After parsing the rules and constructing the call-graph with the *SOOT* framework [22], all tools search for initial analysis seeds. For the creation of the initial analysis seeds, we observed that *SUBS* requires on average 107 milliseconds more time. While we did not adapt the implementation for this step for *CogniCrypt$_{SUBS}$*, we changed the rule to enable *BET*. For collecting the initial seeds, the algorithm collects all rule specifications. Therefore, we attribute this runtime overhead of *SUBS* to the extension of the *CrySL* rules.

After the initialization of seeds, the tools start the actual static analysis to identify misuses. Since we adapted this part, we expect a runtime overhead of *CogniCrypt$_{SUBS}$* compared to *CogniCrypt$_{SAST}$*. Since we adapted this part to implement the detection of error chains, we expect a runtime overhead of *CogniCrypt$_{SUBS}$* compared to *CogniCrypt$_{SAST}$*. The analysis time for *SUBSOFF* and *SAST* have nearly the same average and median time. The subsequent error detection in *SUBSON* propagates hidden predicates and causes a median runtime overhead of 8.4 % for the static seed analysis phase compared to *SUBSOFF*. The improved rule set and BET of *SUBS* add another runtime overhead of 0.4 % compared to *SUBSON*.

> **RQ2**
>
> The subsequent error detection causes a median runtime overhead of 1 % and with BET of 4 %.

## C. Expert Interview (RQ3)

In the previous research questions, we focused on subsequent errors that exist in the wild and the runtime overhead caused by tracking subsequent errors. Our results suggest that subsequent errors exist in the wild, and the runtime overhead for their detection is low. Thus, we were interested in understanding if information about subsequent errors has more usefulness. Besides empirical analyses, reports may also be interesting for users of an analysis. Specifically, we were interested whether grouping and matching subsequent errors to root errors can lead to an improved user experience.

*1) Methodology:* For this research question, we performed an expert interview with a prototype we built that presents root errors and matches subsequent errors grouped below the root error. The prototype differs from the current implementation of *CogniCrypt$_{SUBS}$*, which only reports references between errors, such that the output is formatted to present a root error and beneath the subsequent errors. During our in the wild study (IV-A), we observed that errors sometimes occur in tree structures. Presenting this information is interesting but goes beyond the question if users may benefit from reports of root errors, so we used a simple presentation. Nonetheless, we believe that future work on the usable presentation of subsequent errors is worth exploring.

We recruited five experts for the expert interview. The experts were doctoral- or post-doctoral researchers that are experienced in four different areas. Two researchers are experts for static analyses, one researcher focuses on software tools and domain-specific languages, another is an expert for distributed programming languages, and the last one focuses on IT-security and distributed systems. Our experts have on average 12.6 years experience with Java, and everyone developed at least one project in Java. Three participants developed several big projects and of these one thinks they are an expert. Of our experts, two hesitated to call them a Java expert, as they missed to keep track of the latest Java features. All our experts had at least one course about cryptography at university and three have deeper knowledge because they either used cryptography before, have an IT-Security degree, or even taught cryptographic concepts. Further, three of our participants have prior experience with the Java standard library for cryptography that is used in our code examples.

For the interviews, we used semi-structured interviews. Specifically, we shortly introduced our research and collected the profile of the participants. After the introduction, we asked each participant to solve two different code examples. For each code example, the participants had either the version of *CogniCrypt$_{SAST}$* that supports subsequent errors or not. Due to the number of participants, we could not draw statistical conclusions and did not mix tasks and tools. Thus, the first task should be solved with *CogniCrypt$_{SAST}$* and the second task with our prototype. For the tasks, we asked the participants to follow the think aloud approach [26], [27], and requested the participants to inspect the reports of the analysis. Thus, we avoided that experienced developers resolve the bugs independent of the analysis. We allowed our participants to ask clarification questions or use the internet. After each task, we asked the participants questions on the usability of the respective analysis tool. In the end, we resolved the differences between both rounds and asked final questions. Overall, the interviews had a duration of 40 to 60 minutes. We provided the questions in our artifact (Section VIII).

We executed the expert interviews with a remote-desktop tool to access a virtual machine with Eclipse as IDE installed. The participants could execute the tool by running a script from the integrated terminal. To review the actions and their think aloud, we asked them in advance for consent to record the screen of the VM and audio. Further, we asked each participant for their explicit confirmation when we used direct quotes to ensure that we preserve the intended meaning.

The cryptographic API misuses of both code examples can be fixed by altering one line of code, and are as similar as possible regarding the overall number of errors and root errors. Both examples cover the transmission of a secret with AES encryption. We created the examples based on our experience in the domain with a focus on tasks that can be solved in a reasonable time. The concrete root errors differ between both examples to avoid a learning effect between both examples. The first example spans over four class files with four root errors within three class files. Due to subsequent errors, $CogniCrypt_{SAST}$ reports nine misuses. The second example spans over three class files that are slightly larger than the class files of the first examples. Overall, the second example has three root errors in two class files resolving in a total of ten misuses. The tasks are in our artifact (Section VIII).

*2) Findings:* For the first code example, the participants started to resolve the reports by $CogniCrypt_{SAST}$ applying two main approaches. Three participants started with the error reports for the main class and used them as an entry point for fixing the example. In contrast, two participants started to resolve the errors in the order that $CogniCrypt_{SAST}$ reported them. One participant commented on each error report before they started resolving the error, and thus realized quickly that some errors must be related somehow. Further, some participants used the call-hierarchy function of *Eclipse* to identify the insecure parameter for subsequent errors. In addition, the participants skipped error messages they did not understand or referred to the exact fix location. During the fixing process, these errors often revealed as subsequent errors. Thus, the skipping tactic resolved to be successful for this task.

Error chains caused confusion of participants when a root error was fixed. In particular, when a participant resolved a root error, the number of misuses decreased by more than one. Due to their experience, the participants realized that the errors must be linked to each other or are not relevant anymore. For example, participant four (E4) said: *"No violations found... huh... I only fixed two bugs and three are gone. That's always cool."*. However, in some cases, the participants did not always execute the analysis after resolving a root error and were confused about a subsequent error. For example, participant five (E5) said: *"It is not clear to me how it detects if the key is generated properly... Maybe it detects this from here [pointing to generateKey() method] then it is potentially already fixed. I need to run the analysis again."*.

For task 2, the chosen approach to solve the misuses were more similar among the participants. This time, all participants started with the first reported error that was in the main class. Therefore, the participants may not have changed their approach to resolve the reports. Overall, all participants ran the analysis tool less often than during the first task. Most participants presumed without any explanation from us that they do not need to address subsequent errors, such as participant one (E1): *"Subsequent errors... aha... so this is the Cause-Effect-Chain I'm guessing. Maybe I'll just ignore those and see what happens"*

The expert interview revealed that reporting subsequent errors to users may have a positive impact on the user experience. All participants agreed that $CogniCrypt_{SAST}$ and our prototype supported them in solving the tasks, such as participant five (E5): *"[...]the first tool without the grouping, without the prioritizing each of the error messages had to be taken into consideration to see if they are relevant just to find they have no real meaning because the problem doesn't occur at this location. With the second tool, I immediately found the right errors. This of course is much more pleasant to use [...]."* Further, all experts agreed that during the second task it was more clear which error should be resolved first. However, one expert credited this to the experience gained during the previous task. All five participants would pick our prototype for a use in a security-related project. In addition, two experts are affirmative that subsequent error detecting makes the process to solve cryptographic misuses easier. The three other experts had mixed feelings, as the change "only" helps to identify the location, while the fix is equally challenging for both tools. Three experts agree, one expert is unsure if the presentation of the tool's output increases the acceptance, and one expert disagrees as they argue that for security no error should be ignored.

> **RQ3**
>
> The expert interviews indicate that reporting *root errors* supports users in fixing the relevant errors first.

## V. RELATED WORK

*a) Dependent Errors:* Previous work on fault localization shed a light on the limitations of the assumption that bugs are locally and are isolated. Multiple studies have indicated that bugs are not localized locally [28], [29], [30], [31]. Specifically, faults can spread over many methods or files [28], [29] and even interfere with each other [30]. Thus, their work confirms that bugs can depend more on each other as usually presented and reported by static analyses.

Lipp et al. [18] discussed that vulnerabilities reports may include the root cause, while static analyses tend to report the manifestation location of a vulnerability. To evaluate static analyzers against vulnerability reports, Lipp et al. [18] analyzed for one application if the root cause and manifestation location are in the same function. Their work focused solely on one subsequent error, and their empirical study focused on one application. In contrast, we consider error chains and evaluated 471 applications in our empirical study of error chains.

Besides empirical studies, Avizienis et al. [15] describe a taxonomy of security errors. The taxonomy describes that a root cause can cause an error that is propagated within a program and between different components. The propagation of errors that may cause a fault or vulnerability is called *chain of threats*. While this work describe the propagation in theory, our work showed that the error propagation can cause a tree in practice rather than a chain for cryptographic misuses.

*b) Cryptographic Misuses in the wild:* The majority of empirical studies of cryptographic misuses in the wild draw an alarming picture of today's software security. For Android applications, studies report that 88 % to 99.7 % of the applications contain at least one misuse [32], [3], [33]. An empirical study draw a similar insight for iOS applications [34]. Another study [35] revealed that even Go programs using an API that avoids some misuses by design [36] struggle with cryptographic misuses. Within all analyzed applications, 83 % had at least one cryptographic misuse [35]. For Python applications, a study revealed that 52 % of the applications using a cryptographic API have a misuse [37]. For Maven, a study revealed that 63 % of the artifacts have at least one cryptographic misuse [3]. In contrast to most studies, we "only" identified that 65 % of the analyzed applications have a misuse. Thus, our results are similar to the one by Krüger et al. [3] for Maven artifacts. However, no previous work inspected the dependencies between cryptographic misuses.

*c) Usability of Security Warnings:* Previous work has shown that a high false positive rate hinders the adaption of a static analysis [10], [11], [13]. Our expert interview revealed that *CogniCrypt$_{SUBS}$* may help to reduce confusing reports. Thus, rendering the report to be unresolved [12] or considered as effective false positives [13]. Further, Johnson et al. [10] reported that poorly presented output has a negative impact on the usability of a tool. However, some participants think that a more intuitive presentation can minimize the effect. Our changes are a step in this direction, according to our experts. Further, the different configurations of *CogniCrypt$_{SUBS}$* allow developers to balance between speed and quality [11].

## VI. Limitations & Future Work

Our approach relies on the static analysis *CogniCrypt$_{SAST}$* and the *CrySL* rule set. Thus, the precision and recall of our approach depends on the accuracy of *CogniCrypt$_{SAST}$*, the expressiveness of *CrySL*, and the used rule set. While we tested the correctness of our implementation with test cases, we acknowledge that a more in-depth analysis will be valuable to assess the accuracy. Further, our analysis focuses on the relevant area of cryptographic API misuses due to the well-maintained *CrySL* rule set. Thus, analyses of further APIs are still missing. Future work, could focus on other APIs and understand error chains for these APIs by leveraging our work. Specifically, due to the separation of the analysis logic and the rules, "only" new *CrySL* rules for the APIs of interest need to be specified. Thus, our work may be transferred to other domains.

Recruiting participants for expert interviews or user studies is challenging. For our expert interview (Section IV-C), we only managed to recruit five experts and two of them may have heard about the initial high-level ideas of this research.

Another limitation of the small number of participants is that all participants had the same task and tool order. Therefore, the findings may have a carryover effect between both tasks. We tried to minimize the effect through the design of the tasks, and yet the interview may have a bias. As a consequence, the interviews do not provide statistical insights. Further, for our interviews, we built a prototype that not yet implemented a structured presentation for error chains in the way as they occur in the wild. We acknowledge that implementing and evaluating a structured presentation and a statistically-relevant user study is an interesting future research topic.

The phenomenon we observed in Figure 2 is influenced by the interconnected architecture of cryptographic APIs in Java. Thus, this observation may not generalize to domains in which API constraints are restricted to one class. Further, for other security-related API misuses, presenting the root error may be relevant, too. Consider a taint analysis that can detect SQL injections. While, a SQL injection manifests at the vulnerable call to the database, the dangerous input is the root cause that may occur at a different location. As the analysis can reason about this input, it should be possible for the analysis to report subsequent errors.

## VII. Conclusion

Many applications contain error chains that were so far unstudied. To study these, we introduced *CogniCrypt$_{SUBS}$* which utilizes an existing static analysis and extends its algorithm to identify dependencies between errors. Specifically, we can detect a root error and all its subsequent errors. With *CogniCrypt$_{SUBS}$* we have shown that for cryptographic API misuses error chains occur frequently in the wild. We further find that dependencies between cryptographic API misuses can be expressed as trees. *CogniCrypt$_{SUBS}$* has achieved a negligible median runtime overhead of 1.16 %. Finally, we conducted an expert interview to evaluate qualitatively and show that a structured presentation of root errors and dependent subsequent errors is helpful to fix misuses and improves the usability of the analysis. Overall, we show that error chains as relevant in the wild, and we provide a solution to detect those that can be used for APIs beyond cryptography.

## VIII. Data Availability

*CogniCrypt$_{SUBS}$* is available online as *CogniCrypt$_{SAST}$* release v3.1.0. The scripts and data that we collected during the empirical study (Section IV-A), the benchmark (Section IV-B), and the design and anonymized data of our user study (Section IV-C) is available online on figshare: https://doi.org/10.6084/m9.figshare.24473197.

## Acknowledgment

We are grateful for the valuable feedback that we received from the reviewers that helped us to improve the paper.

This paper is based on work funded by the Deutsche Forschungsgemeinschaft (DFG, German Research Foundation) – SFB 1119 – 236615297, and by the German Federal Ministry of Education and Research together with the Hessian State Ministry for Higher Education (ATHENE).